\magnification=1200
\input amstex.tex
\input amsppt.sty 
%\input arting.sty  
%\input geo.mac 
%\nologo  
\NoBlackBoxes  
\TagsOnRight  
%\pageno=  
\openup 4pt 
\pagewidth{16.5truecm} 
% Largura  
\pageheight{23.00truecm} % Altura  
\def\G{{\bf G}}

\def\I{{\Cal I}}
\def\J{{\Cal J}}

\def\D{{\Cal D}}

\def\e{{\epsilon}}
\def\x{{\xi}}
\def\pa{{\partial}}
\def\s{{\sigma}}
\def\d{{\delta}}
\def\g{{\gamma}}
\def\c{{\chi}}

\def\l{{\lambda}}
\def\Ga{{\Gamma}}
\def\H{{\Cal H}}

%\errorcontextlines=0
%\renewcommand{\rm}{\normalshape}
%redefining \rm to mean: change to roman style
%\DeclareMathOperator{\argmin}{argmin}
%\DeclareMathOperator{\grad}{grad}
%\leftheadtext{ORIZON}
%\rightheadtext{VARIEDADES}
%\begin{document}
$$\eqno{{\text{}} \;\; {\text{}}}
$$
\bigskip
\centerline{\bf BEREZIN INTEGRALS AND POISSON PROCESSES}%\end{center}
\vglue .5in
\centerline{G.F. De Angelis\footnote{ Dipartimento di Fisica, Universit\`a  di 
Lecce, via Arnesano, 73100 Lecce, Italy}; \; G. Jona-Lasinio\footnote
{ Dipartimento di Fisica, Universit\`a di Roma 
 `La Sapienza', Piazzale A. Moro 2, 00185 Roma, Italy}; \;  V. Sidoravicius\footnote{IMPA,
Estr. Dona Castorina 110,  Rio de Janeiro, Brazil and Institute of Mathematics and Informatics,
Akademijos 4, Vilnius 2016, Lithuania}.}
\bigskip
\bigskip
{\sl Dedicated to the memory of Michel Sirugue}
\bigskip
\bigskip 
{\bf Abstract.}
We show that the calculation of Berezin integrals over  anticommuting
variables can be reduced to the evaluation of expectations of functionals
of Poisson processes via an appropriate Feynman-Kac formula.
In this way 
the tools of ordinary analysis can be applied to Berezin integrals
and, as an example, we prove a simple upper bound.  
Possible applications of our results are briefly mentioned.
\bigskip
{\bf{1. Introduction}}
During the last decades the functional integral has become 
the standard approach to the quantization of systems with infinitely many
degrees of freedom. In typical cases like QED and QCD which involve both
bosons and fermions the integral has to deal with anticommuting variables
belonging to a Grassmann algebra and a lucid exposition of the rules
of integration over these variables can be found in \cite{B1}, \cite{FS}.
 
With the discovery of supersymmetry (SUSY) anticommutative integration
has received further impetus and has been applied in different areas
of physics and mathematics.
SUSY  has been first introduced in particle physics 
to express a possible fundamental symmetry between bosons and fermions 
and then has found 
several applications as a formal tool in the theory of complex systems
like heavy nuclei and more recently  disordered
or chaotic mesoscopic systems \cite{VWZ, F, Z}. 
In mathematics it plays a role in various
approaches to index theorems and other topics of differential geometry
 \cite{BZ}.

In physical applications of SUSY one introduces fields $\Psi$
which have two types of components, boson-like $\varphi_1,..., \varphi_{2n}$
which are just ordinary numerical variables and fermi-like
$\xi_1,...., \xi_{2n}$
where the $\xi$-s are anticommuting Grassmann variables.
Supersymmetry transformations
mix boson and fermion components. 

One is usually interested  in the calculation of formal expressions like
$$
\int{d[\phi]d[\xi]\prod_i\varphi_i\prod_j\xi_jexp(-S(\Psi))}
$$ 
called correlation functions. $S(\Psi)$ is a functional invariant
under supersymmetry transformations.

Integrals over anticommuting variables were introduced by Berezin also
in statistical mechanics to represent the partition function of the 
planar Ising model \cite{B2} and the generating function of related
combinatorial problems \cite{B3}. See also \cite{RZ}.
  
The rules of calculation of these integrals,  known as
Berezin integrals, are recalled in the
next section. In spite of the fact that this formalism allows 
compact and very powerful manipulations of the expressions
of interest and in some cases exact calculations,  it
has the drawback that the usual tools of analysis are not
applicable to the anticommuting variables. In particular
one cannot easily find bounds for anticommutative integrals.

The purpose of this
paper is to show that the any  Berezin integral can be represented in terms 
of the expectations of appropriate
functionals of Poisson processes. On the basis of this representation ordinary
analysis can be used and in particular 
upper bounds can be obtained. Furthermore
correlation functions as above can be expressed entirely
as expectations over ordinary stochastic variables.

The starting point of our analysis is a generalized Feynman-Kac
formula developed in the early eighties to express the solutions
of the imaginary time Pauli equation \cite{DJLS}.  The equation is
$$
\partial_t{\psi_t}=-\frac{1}{2}(-i\nabla- \bold{A}(x))^2\psi_t - V(x)\psi_t 
+\frac{1}{2} \bold{H}(x) \cdot {\pmb\sigma}\psi_t  \eqno{(1.1)}
$$
$\bold{A}$ and $\bold{H}$ are the vector potential and the magnetic
field respectively. ${\pmb\sigma}$ denotes the Pauli matrices in the
usual representation. In \cite{DJLS} we proved that the initial value problem is solved by
$$
\psi_t(x, \sigma)=e^t\,\text{\bf{E}}\big[\psi_0(x+w_t, (-)^{N_t}
\sigma)\exp\big{(}-\int_0^tV(x+w_{\tau})d\tau -i\int_0^t
\bold{A}(x+w_{\tau}) \cdot d\bold{w}_{\tau} +
$$
$$
\eqno{(1.2)}
$$
$$
\frac{1}{2}\int_0^tH_z(x+w_{\tau})(-)^{N_{\tau}}\sigma{d\tau} +
\int_0^t\log[\frac{1}{2}(H_x(x+w_{\tau})-i(-)^{N_{\tau}}\sigma{H_y(x+w_{\tau})})]dN_{\tau}\big{)}\big{]}
$$
$\sigma$ is a dichotomic variable which can take the values $\pm 1$.
The expectation is taken with respect to the Wiener process $w_t$
and to the Poisson process $N_t$.
For the understanding of this formula we have to explain
the meaning of the stochastic integral $\int{dN_t}$.
A Poisson process is a jump process characterized by the following
probabilities
$$
P(N_{t+\Delta{t}}-N_t=k)={{(\Delta{t})^k}\over{k!}}e^{-\Delta{t}}
$$
Its trajectories are therefore piecewise constant increasing 
functions and we shall
assume that at each jump they are continuous from the left.
The stochastic integral is just an ordinary Stieltjes integral
$$
\int_0^tf(\tau)dN_{\tau}=\sum_1^nf(\tau_i)
$$
where $\tau_i$ are  random jump times in the interval
$[0,t)$ which are distributed exponentially, that is
$P(\tau\leq{t})=1-e^{-t}$.

The interesting property of formula  (1.2) is that by letting
the spinor indices to become a stochastic process
the Pauli matrices have disappeared from the expression of the 
evolution operator and their algebra is taken into account completely
by the expectation over the jump process. 
The power of the approach was demonstrated by proving
a non trivial paramagnetic inequality which shows that in three dimensions
the evolution is bounded above by the evolution in a magnetic field
which lies in a plane and whose components are simply related
to the original magnetic field. This is easily seen by
taking absolute values and implies for the
ground states
$$
E_0(0, 0, 0; (H_1^2+H_2^2)^{1/2}, 0, H_3) \leq E_0(A_1, A_2, A_3; H_1, H_2, H_3)
$$

For a recent application of the \cite{DJLS} approach see \cite{ER}.
 
Since the Pauli matrices are objects which belong to a Clifford algebra
the above findings suggested a possible connection between calculus
with Poisson processes and calculus with anticommuting variables.  
It is the main purpose of this paper to implement such a connection.
To illustrate how the connection comes about in Sec. 3 we take up again 
the case of 
Pauli type equations and we observe that they can be interpreted as evolution
equations over a Grassmann algebra. This type of evolution was considered
for instance by Berezin and Marinov \cite{BM}, \cite{M}.   
Solutions of evolution equations over Grassmann algebras can be expressed
as Berezin integrals which represent the convolution  of the kernel
of the evolution operator with the initial condition. A straightforward 
comparison with the solution given in \cite{DJLS} provides the
identification of the anticommutative integrals with the 
appropriate Poisson expectations.

In order to develop the theory in a systematic way  for an arbitrary but finite number of anticommutative variables in Sec. 4 we introduce
a representation of Grassmann algebras in the space of functions of
dichotomic variables that we call the $\sigma$-representation. This
space was used by Wigner in his book on group theory \cite{W}
to find the representations of the symmetric group connected with
the exclusion principle.   

In Sec. 5 we develop the necessary theory of semigroups associated
with Poisson processes and in Sec. 6 we make the identification
of Berezin integrals with appropriate expectations. We also derive 
a general inequality.

In Sec. 7 we discuss in particular the Gaussian anticommutative  integrals
due to their importance in physical applications. 
   
We conclude this introduction with some comments on 
possible interesting applications of the results obtained in this paper.
 The semigroups associated to Poisson processes 
 encompass all Hamiltonian semigroups 
${\{ {\text {exp}}(-tH)\}}_{t\geq 0}$ for $k$ interacting $\frac{1}{2}$-spins in an external magnetic field as,
for instance, a Heisenberg ferromagnet or, by a slight change of
language, any Hamiltonian semigroup for models describing interacting fermions on a finite lattice.
Our representation can therefore be used both for theoretical or simulation studies of the statistical mechanics of such systems. 

A particularly interesting case to which our approach can be applied is the
calculation of the Dirac propagator on a lattice, an important
problem in the study of QCD on the lattice. Presumably in this way
it is possible to obtain a  simplification of the methods
used at present in simulations. See the remark at the end of section 7.

More generally in all cases where SUSY is relevant our approach
may be a useful tool.

We believe that our results have also an independent interest
in so far as they establish a direct connection between algebraic
objects as those represented by Berezin integrals and
analytic expressions. 

As a final remark we observe  that the Wiener process and the
Poisson process are both Levy processes and are actually the
two limit cases of the Levy-Khinchin formula. We find therefore
quite satisfying their correspondence at the euclidean level with the two basic types
of particles in nature, bosons and fermions. Then the question
naturally arises: do the other
processes described by the Levy-Khinchin formula have any
relevance for physics?

\bigskip
{\bf{2. Analysis on Grassmann Algebras}}
\bigskip
We open this Section by a short review of standard definitions and results 
and refer reader to [B], [D], [FS] for more detailed information.

By ${\text{\bf{G}}}(k)$ we denote the Grassmann algebra over ${\Bbb {C}}$ generated by its identity {\bf{1}}$_k$ and a family $\{\xi_1,\ldots,\xi_k\}$ of
generators which obey the following commutation relations:
$$
 \xi_i\xi_j = -\xi_j\xi_i\,; \quad \forall\,i,j. \eqno{(2.1)}
$$

For future reference, {\bf{G}}$^n (k)$ $\simeq$ {\bf{G}}$({nk})$ will be the Grassmann algebra 
generated by {\bf{1}}$_k^n$ and $\{\xi_1^1,\ldots,\xi_k^1,$ $\xi_1^2,\ldots,\xi_k^2\,, $ $
\dots \,,\xi_1^n,\ldots,\xi_k^n\, \}$,
and by convention, {\bf{G}}$^1 (k)$ = ${\text{\bf{G}}}(k)$.

The collection $\{1,\dots ,\, k\, \} $ of labels is any non - empty finite set endowed with a
total ordering.

Elements of {\bf{G}}(k) of the form
$\x_{i_1}\x_{i_2}\cdot\ldots\cdot\x_{i_n}$ are called monomials; we will
use the set of ordered multiindices $M_k = \{\mu = (\mu_1,\ldots,\mu_n)\,:\,
1 \leq \mu_1 < \mu_2 <\ldots < \mu_n\  \leq  k\}$, and write:
$$
\x^\mu = \x_{\mu_1} \cdot\dots\cdot \x_{\mu_n}\,,
$$
   for $\mu = (\mu_1,\mu_2,\ldots,\mu_n)$.
As a linear space {\bf{G}}(k) has dimension $2^k$, and each element  $F(\x) \, \in$  {\bf{G}}(k) 
can be represented in a unique way as a polynomial with complex coefficients:
$$
F(\x) = F(\x) = \, f_0 \cdot {\text{\bf{1}}}_k + 
\sum_{r=1}^k \, \sum_{1\leq i_1 <\dots\, < \, i_r \, \leq \, k} 
f_{i_1,\dots,i_k}\x_{i_1}\cdot \dots \cdot \x_{i_r} \, = \, 
\sum_{\mu\in M_n} f_\mu\cdot \x^\mu\,, \eqno{(2.2)}
$$
where $f_\mu \in \, {\Bbb{C}}$, and
 therefore {\bf{G}}$(k)$ is naturally graded. It is advantageous thinking of $F(\x)$ as a ``function''
of the Grassmann ``variables'' $\x_1,\,\dots \,,\x_k$ that is to say of the ``Fermi field''  $\{\x_1,\ldots,\x_k\}$.

Analysis over {\G}$(k)$ is based upon left derivatives  $\frac{\d}{\delta \x_1},
\dots, \,\frac{\delta}{\delta \x_k}$ and Berezin integral, which are defined as follows:
$$
\frac{\delta}{\delta \x_i} \x_{\mu_1} \cdot\dots\cdot \x_{\mu_n}\, \overset{\text{def}}\to{=} \,
\d_{\mu_1 i} \cdot  \x_{\mu_2} \cdot\dots\cdot \x_{\mu_n}\,-\,
\d_{\mu_2 i} \cdot  \x_{\mu_1} \x_{\mu_3}\cdot\dots\cdot \x_{\mu_n}\,+\,
{(-1)}^{k-1}\d_{\mu_k i} \cdot  \x_{\mu_1} \cdot\dots\cdot \x_{\mu_{k-1}}\,=
$$
$$
\sum_{j=1}^k {(-1)}^{j-1}\d_{\mu_j i} \cdot  \x_{\mu_1} \cdot\ldots 
{{\x\!\!\!/_{\mu_j}}}\cdot\dots \cdot
 \x_{\mu_{k-1}}\,,
$$
where the sign $/$ over the generator $\x_{\mu_j}$ means that it is omitted. 

To define an integral we introduce symbols $d\x_1,\ldots,d\x_k$ satisfying the
following commutation relations:
$$
\{d\xi_i, d\xi_j\} = \{d\xi_i, \xi_j\} = 0;
$$
where $\{a,b\} = a\cdot b + b\cdot a$, and define ``basic'' integrals:
$$
\int^B d\xi_i = 0, \qquad \int^B \xi_\mu\cdot \xi_i\,d\xi_i = \xi_\mu\,;
$$
if $\mu_j \ne i$ for all $j$ in $\mu$. We extend the integral on
{\bf {G}}(k) by linearity and call it the Berezin integral.  
In general
$$
\int^B F(\x) d\x_k\cdot \dots \cdot d\x_1\, \overset{\text{def}}\to{=} \,
\int^B F(\x) {\D}_k \x \, \equiv \, f_{1,\dots,k} \, \in {\Bbb{C}} \,,  \eqno{(2.3)}
$$
and observe that up to a multiplicative constant, is uniquely defined as the only linear form over
{\G}(k) alternating under permutations of the Grassmannian variables. It transforms [B]
as
$$
\int^B F(R\x) {\D}_k \x \, = \, ({\text{det}}R)\cdot\int^B F(\x)
 {\D}_k \x \,, \eqno{(2.4)}
$$
under the linear substitution $\x_i \, \to \, \sum_{j=1}^k {R}_{ij}\x_j$.

A specially important case is provided by ``Gaussian integrals'':
$$
\int^B {\text{ exp}} \{ \frac{1}{2}\sum_{i,j=1}^k {A}_{ij} \x_i\x_j\} 
{\D}_k \x\, ;
\eqno{(2.5)}
$$
where one can always assume that {{A}} = ({{A}}$_{ij}$) is an antisymmetric matrix, otherwise it
could be replaced by $2^{-1}$({{A}} - {{A}}$^T$). ``Gaussian integrals'' vanish for odd k
while, by exploiting (2.4), we get
$$
\int^B{\text{ exp}} \{ \sum_{1\leq h < k\leq 2k} {A}_{hk} \x_h\x_k\} {\D}_{2k} \x\,
=\, Pf \, {A}\, ;
\eqno{(2.6)}
$$
The Pfaffian $Pf$ {{A}} of the triangular array ${\{{A}_{hk}\}}_{1 \leq h <k\leq 2k}$
is defined by :
$$
Pf A \, = \, \sum_{\pi} \, (-1)^{\pi}{A}_{i_1 j_1} \, \cdot \dots \cdot
{A}_{i_k j_k}\,; \eqno{(2.7)}
$$
where the sum $\sum_{\pi}$ is taken over all $(2k)!/2^k k!$ ways of pairing of the elements
of the set $\{1,\,2,\dots,
2k\}$ where $(-1)^\pi$ is the parity of the permutation $\pi \, = \, (i_1 j_1,\dots , i_k j_k)$.

Let us consider the operator of left derivative $\frac{\delta}{\delta\xi_i}$  and the operator 
$\widehat{\xi}_i$ of left multiplication by
the element $\xi_i$, both acting on {\bf {G}}(k) (see [B]). In the sequel we shall omit the hat if confusion does not arise.

We recall that all linear operators acting on {\bf {G}}$(k)$ belong to the Clifford 
(or Spinor) algebra {\bf {C}}(2k)  generated by its identity ${\hat {\text{\bf {1}}}}_{2k}$
and the operators  $\frac {\delta}{\delta \x_1}, \dots,\frac {\delta}{\delta \x_k}$, and
${{\x}}_1,\dots,{ {\x}}_k$, which satisfy the following commutation relations:
$$
\bigg\{ { {\x}}_i,\, \frac {\delta}{\delta \x_j} \bigg\} \, = \, \d_{ij}\, ,\qquad 
\bigg\{ { {\x}}_i,\, {\x}_j \bigg\}\, = \,
\bigg\{ \frac {\delta}{\delta \x_i}, \, \frac {\delta}{\delta \x_j} \bigg\} \, = \, 0\,,
\qquad i,j \, = \,
1, \dots, k.
$$
 The algebra {\bf {C}}(2k) 
is isomorphic to the CAR algebra [BR] over a $2^k$ - dimensional Hilbert space since operators
$\frac {\delta}{\delta \x_i}$ and ${ {\x}}_i$ might be interpreted as annihilation and
creation operators for a Fermi system with $k$ degrees of freedom.

We end the Section by defining kernels of operators acting on {\bf {G}}(k). We recall that to
each $L \, \in$ {\bf {C}}(2k) corresponds a unique element {\text{Ker}}$(L)(\xi,\x')$ of 
the Grassmann algebra {\bf{G}}(2k), generated by $\x_1, \dots , \x_k$, and ${\x'}_1 , 
\dots, {\x'}_k$, such that
$$
(LF)(\x)\, = \, \int_B {\text{Ker}}(L)(\xi,\x') \cdot F(\x') \,\,{\D}_k\x' \, , 
$$
%.....................................................................................
\bigskip
\bigskip
{\bf{3. Evolution on G($k$)}}
\bigskip
Let us consider the evolution given by the equation:
$$
{\frac{\pa f_t}{\pa t}} = L\,f_t \, ;\eqno{(3.1)}
$$
where $f_t \in$ {\bf{G}}$(k)$ and $L \in$ {\bf{C}}(2k).

In this section first we solve the equation (3.1) by constructing a kernel for the
operator $\exp(tL)$ using standard tools of Grassmannian analysis and writing the solution of (3.1)
as  an element of {\bf{G}}$(k)$ with coefficients given by certain Berezin integrals, which we
compute explicitely, and second, we show that these integrals could be represented as 
expectations with respect to a properly chosen family of standard Poisson processes.

We introduce some necessary constructions related to the formal 
description of continuous time
evolution on {\bf{G}}$(k)$. In order to do that it is
convenient to embed {\bf{G}}(k) into an extended Grassmann algebra
{\bf{G}}$_\infty(k)$, and we will proceed in the following way (cf 
[ MIS]). Consider three Grassmann algebras:

the Grassmann algebra ${\Gamma}_{\tau} (k)$, $\tau = (t_1,\ldots,t_m)$,
generated by:

\centerline{{\bf{1}}, $\x_1(t_i),\ldots,\x_k(t_i); \; \rho_1(t_i),\ldots,\rho_k(t_i);\qquad 
i=1,\ldots,m;$}%\end{center}

the Grassmann algebra ${{\Gamma}}$$_{t]}(k)$ generated by:

\centerline{{\bf{1}}, $\x_1(s),\ldots,\x_k(s);\; \rho_1(s),\ldots,\rho_k(s); \qquad 
s \le t;$}%\end{center}

and the Grassmann algebra ${{\Gamma}}$$_\infty(k)$ generated by:

\centerline{{\bf{1}}, $\x_1(s),\ldots,\x_k(s);\; \rho_1(s),\ldots,\rho_k(s);  \qquad s > 0;$}%\end{center}
\noindent where 
$$
\x_i(s) \cdot \x_j(t)\, = \, - \,\x_j(t) \cdot \x_i(s);\; 
\rho_i(s)\cdot \rho_j(t)\, = \, -\, \rho_j(t) \cdot \rho_i(s),\;
$$
and
$$
\rho_i(s)\cdot \x_j(t)\, = \, - \,\x_j(t) \cdot \rho_i(s) \quad \forall\,i,j,\; \; 
\forall\,s,t > 0. 
$$

We consider exterior algebras:

\centerline{{\bf{G}}$_{\tau}(k)$ ={\bf{G}}$_0 (k)$ $\bigwedge $ ${{\Gamma}}$$_{\tau}(k)$; }%\end{center}

\centerline{ {\bf{G}}$_{t\, ]} (k)$
= {\bf{G}}$_0 (k)$ $\bigwedge$ ${{\Gamma}}$$_{t\, ]}(k)$; }%\end{center}

\centerline{ {\bf{G}}$_{\infty} (k)$ = {\bf{G}}$_0 (k)$ $\bigwedge$
 ${{\Gamma}}$$_{\infty} (k)$; }%\end{center}
\noindent where {\bf{G}}$_0(k)$ = {\bf{G}}(k).

{\bf {Remark.}} a) For the construction of Grassmann algebras with an infinite number 
of generators we refer also to [B], [R], [S].  b) The generators $\rho$'s 
will be used only for  
"Fourier transform" - type expressions and play only an auxilliary role here.
\bigskip

\noindent {\bf {Example.}}  The Pauli equation on {\bf{G}}(1).

\noindent Let us consider evolution on {\bf{G}}(1), 
 generated by {\bf{1}} and $\xi_1$, given by
(3.1) where
$$
L = h_1\,\sigma_1 + h_2\,\sigma_2 + h_3\,\sigma_3\,;
$$
and we identify:
$$
\sigma_1 = {\xi} + \frac{\d}{\d\xi}\,; \qquad \sigma_2 =
 i \left({\xi} - {\frac{\d}{\d\xi}}\right); \qquad
\sigma_3 = \frac{1}{i} \s_1 \s_2\,;
$$
so that the usual commutation rules of Pauli matrices are satisfied.

As it will be proven in Theorem (3.1) the kernel of the operator $exp(tL)$ can be written as the limit:
$$
{\text{Ker}}(e^{tL})(\xi,\x') = \lim_{m\to\infty}\,Q_t^m(\xi,\x');
$$
where
$$
Q_t^m(\xi,\x') = \underbrace{\int^B \cdot\ldots\cdot\int^B}_{m-\text{times}}
P_t^m(\xi,{\x_t},\rho_t,\x')\D_m\rho_t\,\D_m{\x_t} \; ;
$$
which is the kernel of the operator ${({\text{\bf{1}}}\, + \, \frac{t}{m} L)}^m$ and $P_t^m(\xi,{\x_t},\rho_t,\x')\, \in$  
{\bf{G}}$_0 (1)$ $\bigwedge \Gamma_{\tau (t,m)}(1)$, with 
$\tau (t,m) \,= \, (t/m,\; 2t/m,\; \dots , [(m-1)t]/m,\; t),$  and has the following form:
$$
P_t^m(\xi,{\x_t},\rho_t,\x')\, =
$$
$$
={\text{exp}} \bigg\{ \sum_{j=0}^m \frac tm \bigg\{ h_1\left[\x\left(\frac{j\cdot t}{m} \right)
+ \rho\left(\frac{j\cdot t}{m}\right)\right] +
ih_2\left[\x\left(\frac{j\cdot t}{m}\right) - \rho\left(\frac{j\cdot t}{m}\right) \right]  \, + 
$$
$$
+ h_3\left[{1} - 2\x\left(\frac{j \cdot t}{m}\right)
\rho\left(\frac{j\cdot t}{m}\right)\right]- \rho\left(\frac{j\cdot t}{m}\right) \left[\x
\left(\frac{j\cdot t}{m}\right) +
\x\left({\frac{(j-1)t}{m}}\right) \right] \bigg\} \bigg\}; 
$$
where 
$$
\D_m\rho_t = d\rho\left(\frac tm\right) \cdot\ldots\cdot d\rho(t);
$$
$$
\D_m{\x_t} = d\x\left(\frac tm\right) \cdot\ldots\cdot d\x\left(\frac{(m-
1)t}{m}\right);
$$

\noindent and we put $\x(0)=\x'$, $\x(t) = \xi$.
\noindent In this way the solution of Equation (3.1) with initial data
$F(\xi) \, = \, f_0 \cdot${\bf{1}} + $ f_{1} \cdot \xi$;
 $ f_0 , \, f_{1} \, \in \,  {\Bbb{C}}$ can be written:
$$
F_t(\xi) = (e^{tL}\,F)(\xi) = \int^B \left[{\text{Ker}}(e^{tL})(\xi,\x')F(\x')\right]d\x'; 
\eqno{(3.2)}
$$
and we have:
$$
F_t(\xi) = f_0  (t){\text{\bf{1}}}\, +\,  f_{1}(t)\cdot \x\, ,
$$ 
where
$$
f_0(t) = -\int^B \int^B \left[{\text{Ker}}(e^{tL})(\xi,\x')F(\x')\cdot\xi \right]\,d\x'\,d\xi \eqno{(3.3)}
$$
$$
f_{1}(t) = \int^B\int^B \left[{\text{Ker}}( e^{tL})(\xi,\x')F(\x') \right] d\x'\,d\xi.\ 
\eqno{(3.4)}
$$
On the other hand from equation (1.2) specialized to the present simplified case we have that:
$$
f_0(t) = e^t\,\text{\bf{E}}\big[f_{{1-(-1)^{N_t}\over{2}}}
$$
$$
\cdot \exp\big(\int_0^t \ell n(h_1 - i(-1)^{N_{\tau}}\,h_2) dN_{\tau}
- \int_0^t h_3(-1)^{N_{\tau}}\,d\tau)\big)\big];  \eqno{(3.5)}
$$
and
$$
 f_{1} = e^t\,\text{\bf{E}}\big[f_{{1+(-1)^{N_t}\over{2}}}
$$
$$
\cdot \exp\big(\int_0^t \ell n(h_1+i(-1)^{N_{\tau}}\,h_2)dN_{\tau} + 
\int_0^t h_3(-1)^{N_{\tau}}\,d\tau\big)\big];  \eqno{(3.6)}
$$

where $N_t$ is a Poisson process with unit parameter. Comparing (3.3) and (3.5) we get:
$$
-\int^B\int^B \left[{\text{Ker}} (e^{tL})(\xi,\x')F(\x')\xi \right] \,d\x'\,d\xi =e^t\,\text{\bf{E}}
\big[f_{{1-(-1)^{N_t}}\over{2}}\cdot
$$
$$
\cdot \exp\big(\int_0^t \ell n(h_1-i(-1)^{N_{\tau}}\,h_2)dN_{\tau} -
\int_0^t h_3(-1)^{N_{\tau}}\,d\tau\big)\big], \eqno{(3.7)}
$$
An analogous equation can be written for $f_1 (t)$.

In this way we obtain the equality between certain Berezin integrals and
expectations with respect to the standard Poisson process.
\bigskip

To discuss the general case, let us consider the
family of operators:
$$
\g_j = {\xi}_j + \frac{\d}{\d \xi_j}\,; \qquad \; \bar{\g}_j =
\frac{1}{i} \left({\xi}_j - \frac{\d}{\d\xi_k}\right), \qquad
j=1,\ldots,n
$$
which satisfy the commutation relations:
$$
\{\g_i, \bar{\g}_j\} = 0, \qquad \{\g_i,\g_j\} = \{\bar{\g}_i,
\bar{\g}_j\} = 2\delta_{ij}\,,
$$

Further we will use the
notations: let$\nu$, $x$, $\mu \in M_k$ 
$$
\nu = \, \{\nu_1,\ldots,\nu_n\}, \quad x = \{x_1,\ldots,x_{\ell}\}, \quad \mu = \{\mu_1,\ldots,\mu_m\},
$$
such that
$$
\nu \cap \mu = \nu \cap x = s \cap \mu = \emptyset;
$$
we will write:
$$
\g^{(\nu,x,\mu)} = \prod_{i=1}^n \g_{\nu_i} \cdot \prod_{j=1}^{\ell} 
(\bar{\g}_{x_j}\,\g_{x_j}) \cdot \prod_{r=1}^m \bar{\g}_{\mu_r}\,. \eqno{(3.8)}
$$
Let us set:
$$
L = \sum_{(\nu,x,\mu)} h_{(\nu,x,\mu)}\,\g^{(\nu,x,\mu)}\,.
$$
where $\g^{(\nu,x,\mu)} \, \in \text{\bf{C}}(2k)$ and 
$h_{(\nu,x,\mu)}\, \in {\Bbb{C}}$.
\bigskip
\noindent {\bf{Theorem 3.1.}} The kernel of the operator $\exp(tL)$ acting on {\bf{G}}(k) is given 
as the limit:
$$
\text{{Ker}}(e^{tL})(\xi,\x') = \lim_{m\to\infty} \,\,Q_t^m(\xi,\x')\, \in \, {\text{\bf{G}}}(2k)\, ;
$$
where
$$
Q_t^m(\xi,\x') =
$$
$$
\int^B \exp\bigg\{-\sum_{j=1}^m \bigg(-\frac{j\cdot t}{m} \sum_{(\nu,x,\mu)}
\bigg[h_{(\nu,x,\mu)} \prod_{\nu} (\xi_\nu + i\rho_\nu) \cdot
$$
$$
\cdot \prod_x (2\xi_x\rho_x + i{\text{\bf{1}}}) \prod_{\mu} \frac 1i (\xi_\mu -
i\rho_\mu)\bigg]\bigg) -
$$
$$
- \sum_j^m \sum_{r=1}^n \rho_r
\left(\frac{j\cdot t}{m}\right)\left[\x_r'\left(\frac{j\cdot t}{m}\right) \, + \,
\x_r'\left(\frac{(j-1)t}{m}\right)\right]\bigg\}
\D\rho\, \;
$$
if k is even, and similar formula (see Appendix 1.) holds for k  odd.

Proof. The proof is rather simple and we postpone it to Appendix 1.

\noindent In this way we can write solution of the equation (3.1) as the Berezin integral:
$$
F_t (\xi)\, =\, (e^{tL} \, F_0)(\xi)\, = \, \int^B [\text{{Ker}}
(e^{tL}) (\xi ,\x' ) F_0 (\x' ) ] \, \D\x'\, =
$$
$$
=\, \sum_{\mu \in M}
f_{\mu}(t)\cdot \xi^{\mu}\, ;
$$
with $f_{\mu} (t)\, \in \, {\Bbb{C}}$.

Using Berezin integration rules we immediately get:
$$
f_{\mu} (t)\, = \int^B \, [ \, F_t (\xi)\, \cdot \xi^{{\mu}^c}] \, \D\xi \, =
$$
$$
=\, \int^B \bigg\{ \int^B [\text{{Ker}}
(e^{tL}) (\xi ,\x' ) F_0 (\x' ) ]  \, \D\x'\,\cdot \xi^{{\mu}^c} \,
\bigg\} \, \D\xi; \eqno{(3.9)}
$$
where $\mu^c\, \in \, M$ is a complementary multiindex to $\mu$, i.e. 
$\mu  = (\mu_1,\ldots,\mu_n)$ and $\mu^c  = (\mu_1^c,\ldots,\mu_{n'}^c)$,
are such that $\{\mu_1,\ldots,\mu_n\}\, \bigcap \, \{\mu_1^c,\ldots,\mu_{n'}^c \} = 
\emptyset$ and  $\{\mu_1,\ldots,\mu_n\}\, \bigcup \,$ $ \{\mu_1^c,\ldots,\mu_{n'}^c \} =
\{1,\, 2, \dots, \, k\}$, and we assume here
that $\emptyset^c \, = \, \{1,2,...,k\}$.

Let us fix a total ordering $\prec$ on M (for example lexicographic order). It
induces a total ordering on the set of monomials  $\xi^{\mu}$  in the following
way:
$$
\xi^{\mu}\, \prec \, \xi^{\mu'} \qquad \; \text{if} \; \mu \, \prec \, \mu'\, ;
$$
let us rename all monomials with respect to the order $\prec$ by
$\xi^1,\, \xi^2, \dots, \xi^{2^k}$, which form the basis of {\bf{G}}$(k)$
as a $2^k$-dimensional linear space.

In this basis the matrix elements of the operator $\g^{(\nu,x,\mu)}\, =
\, \prod_{i=1}^n \g_{\nu_i} \cdot \prod_{j=1}^{\ell} 
(\bar{\g}_{x_j}\,\g_{x_j}) \cdot \prod_{r=1}^m \bar{\g}_{\mu_r}$ are explicitely computable, 
so equation (3.1) could be
rewritten:
$$
{\frac{\pa {f}_{\alpha}(t)}{\pa t}} = \sum_{\beta\, = \, 1}^{2^k} 
{L}_{\alpha, 
\, \beta} \,  {f}_{\beta}(t) \, ;\eqno{(3.10)}
$$
where the coefficients ${L}_{\alpha, \, \beta}\, \equiv \, 
{L}_{\alpha, \, \beta}( h_1,\, h_2, \dots, h_{2^k})\, 
\in \,{\Bbb{C}}$.

With an obvious definition of $\Phi (\alpha, \beta )$ and $\Psi (\alpha)$, equation
(3.10) can be rewritten
$$
{\frac{\pa {f}_{\alpha}(t)}{\pa t}} = \sum_{\beta\, = \, 1}^{2^k -\, 1} 
{\text{exp}}[\, \Phi (\alpha, \beta )]\cdot  \,   {f}_{\alpha \oplus \beta}(t) \, +
 \, \Psi (\alpha) \cdot  {f}_{\alpha}(t);\eqno{(3.11)}
$$
with the initial condition
$$
 {f}_{\alpha}(0)\, = \,  {f}_{\alpha};
$$
and where the sign $\oplus$ stands for the sum modulo $2^k$.

The solution of the linear system (3.11) is given by
$$
{f}_{\alpha}(t)\, = \, e^{(2^k -\, 1)\cdot t} \cdot \text{\bf{E}}
\bigg[{f}_{\alpha \, \oplus \, N_t} \cdot {\text{exp}} \bigg( \int_0^t \,  
\Psi (\alpha \, \oplus \, N_t\, \ominus \, N_{\tau})\, d\tau \; +
$$
$$
+\;  \sum_{\beta\, = \, 1}^{2^k -\, 1} \,  \int_0^t \, \Phi (\alpha\, \oplus\, N_t\, \ominus \, N_{\tau}\, \ominus \, \beta , \, \beta )\, d\, N_{\tau}^{\beta} 
\bigg) \bigg]\, ; \eqno{(3.12)}
$$
where $N_t \, = \, \sum_{\beta \, =\, 1}^{2^k \, - \, 1} \beta N_t^{\beta}$ is the 
sum of $2^k \, - \, 1$ independent Poisson processes and  $\alpha\, \oplus\, N_t , \; \alpha \, \oplus \, N_t\, \ominus \, N_{\tau}$, and
$\alpha\, \oplus\, N_t\, \ominus \, N_{\tau}\, \ominus \, \beta$ stand for  sums
and differences modulo $2^k$. (For complete proof see [DJLS].)

Now comparing (3.9) and (3.12)  
$$
\int^B \bigg\{ \int^B [\text{{Ker}}
(e^{tL}) (\xi ,\x' ) F_0 (\x' ) ]  \, \D_k\x'\,\cdot \xi^{{\mu}^c} \,
\bigg\} \, \D_k\xi\, = 
$$
$$
\, e^{(2^k -\, 1)\cdot t} \cdot \text{\bf{E}}
\bigg[{f}_{\alpha \, \oplus \, N_t} \cdot exp \bigg( \int_0^t \,  \Psi (\alpha \, \oplus \, N_t\, \ominus \, N_{\tau})\, d\tau \; +
$$
$$
+\;  \sum_{\beta\, = \, 1}^{2^k -\, 1} \,  \int_0^t \, \Phi (\alpha\, \oplus\, N_t\, \ominus \, N_{\tau}\, \ominus \, \beta , \, \beta )\, d\, N_{\tau}^{\beta} 
\bigg) \bigg]\, ; \eqno{(3.13)}
$$
Sums and differences modulo $2^k$ are slightly unconfortable
to handle. In the next three sections we shall reformulate the theory in a space of functions of dichotomic variables. This allows
the construction of a systematic formalism
suitable for both theoretical and numerical analysis.

\bigskip
\bigskip
{\bf {4. Representation of Grassmann Algebras in the Space of Functions of Dichotomic Variables
($\sigma$ - Representation)}}
\bigskip

In this Section we discuss a linear bijection between Grassmann algebras and spaces of
functions of dichotomic variables.

Let 
{\bf{Z}}$_2$ be $\{ -1, \, 1 \}$ with its natural Abelian group structure and {\bf{Z}}$_2^
{\times k}$ be the direct product of $k$ copies of {\bf{Z}}$_2$ which is finite commutative 
group with the unit element $e_k \, = \, (1, \, \dots , 1)$.

Let us define ${\Cal{H}}_k$ as the linear space of all ``wave functions'' $\c (\cdot)$ :
 {\bf{Z}}$_2^{\times k} \, \to \, {\Bbb {C}}$ of $k$ dichotomic variables $\sigma_1,\dots,\, \sigma_k$.
It becomes a $2^k$ - dimensional Hilbert space when equipped with the inner product:
$$
{< \c_{_1}\, , \c_{_2} >}_k \, = \, 
\sum_{\sigma \in {\text{{\bf{Z}}}}_2^{\times k}}\, {\bar{\c}}_{_1} (\sigma) \c_{_2} (\sigma)\, ;
 \eqno{(4.1)}
$$
and could be interpreted as the space of (pure) states for a Heisenberg ferromagnet in a finite
box. 

All monomials of the Grassmann algebra {\bf{G}}$(k)$ can be indexed by elements of
${\text{{\bf{Z}}}}_2^{\times k}$ in the following way:
$$
\s \, \mapsto \, \x (\s) \, \equiv \, \x^{\frac{1-\sigma}{2}}\, = \, \x_1^{\frac{1-\sigma_1}{2}}
\, \dots \,\x_k^{\frac{1-\sigma_k}
{2}}\, , \qquad \sigma\, = \, ( \sigma_1,
\, \dots \, , \sigma_k) \, \in {\text{{\bf{Z}}}}_2^{\times k}, \eqno{(4.2)}
$$
and let us define now a map $\I$: $\H_k \, \to$ {\bf{G}}$(k)$ by the formula:
$$
\I(\c) \, \equiv \, F_{\c}(\x) \, = \, \sum_{\sigma \in {\text{{\bf{Z}}}}_2^{\times k}}
\,\c(\sigma) \cdot \x^{\frac{1-\sigma}{2}}\,\in \, {\text{\bf{G}}}(k) \,; \eqno{(4.3)}
$$
where the ``wave function'' $\c(.)\, \in \, {\H}_k$.

It is easy to see that the map $\I$ is injective, i.e. $\I(\c_{_1}) \, \neq \, \I(\c_{_2})$ if 
$\c_{_1} \, \neq
\, \c_{_2}$. Moreover, the map $\J \,: \;{\text{\bf{G}}}(k) \, \to \, \H_k $ defined by:
$$
\J (F(\x))\, \equiv \, \c_{_{F}}(\sigma)\, = \, 
\triangle_k (\sigma) \int^B \x^{\frac{1+\sigma}{2}}\cdot F(\x){\D}_k \x \,,
\eqno{(4.4)}
$$
where  $\triangle_k (.) \, : \, {\text{{\bf{Z}}}}_2^{\times k}\; \to \; {\text{{\bf{Z}}}}_2$ is 
given by
$$
\triangle_k (\sigma)\, =\, \prod_{l=1}^k \bigg( \frac{1-\sigma_l}{2} \, +\, \frac{1+\sigma_l}{2}
\cdot \sigma_1 \cdots \sigma_{l-1}\bigg)\, . \eqno{(4.5)}
$$
is the inverse of   $\I$  :  $\I \, = \, {\J}^{-1}$, and, clearly, it is surjective. By this we 
obtain a linear bijection between Grassmann algebra ${\text{\bf{G}}}(k)$ and the space of 
functions of dichotomic variables ${\H}_k$, which we call $\s$ - representation.

Next we turn to {\bf{C}}${(2k)}$. To each ${\widehat {A}} \, \in \, {\text{\bf{C}}}(2k)$ 
corresponds a linear operator $A \, : \, {\H}_k \, \to \, {\H}_k$ given by formula:
$$
{\widehat A} \, F(\x)\, = \, \sum_{\sigma \in {\text{{\bf{Z}}}}_2^{\times k}} (A \c)(\sigma)
\cdot \x^{\frac{1-\sigma}{2}}\, . \eqno{(4.6)}
$$

For any $\widehat A \, \in \, {\text{\bf{C}}}(2k)$, which is a linear combination of normally 
ordered products ${ {\x}}^{\frac{1-\epsilon}{2}}\cdot {(\frac{\delta}{\delta \x})}
^{\frac{1-\eta}{2}}$, where $\epsilon ,\eta \, \in$ {{\bf{Z}}}$_2^{\times k}$, we find its image
 $A$ by computing the image of operators ${{\x}}^{\frac{1-\epsilon}{2}}\cdot {(\frac{\delta}{\delta \x})
^{\frac{1-\eta}{2}}}$ which we denote by $a^{*\frac{1-\epsilon}{2}}a^{\frac{1-\eta}{2}}$. 

\noindent {\bf {Proposition 4.1}} For all $\epsilon , \, \eta \, \in$ {{\bf{Z}}}$_2^{\times k}$ we 
have:
$$
(a^{*\frac{1-\epsilon}{2}} \c)(\sigma) \, = \, C_k (\epsilon, \sigma) \cdot 
\c(\epsilon \sigma)\,,
\qquad (a^{\frac{1-\eta}{2}} \c)(\sigma)\, = \, A_k (\eta, \sigma) \cdot \c(\eta \sigma)\,,
$$
$$
(a^{*\frac{1-\epsilon}{2}} 
a^{\frac{1-\eta}{2}} \c)(\sigma)\, = \, N_k (\epsilon,\eta, \sigma) \cdot
 \c(\epsilon \eta \sigma)\,;
$$
where
$$
C_k (\epsilon, \sigma)\, = \, \prod_{l=1}^k \bigg( \frac{1+{\epsilon}_l}{2}\, +\,
\frac{1-{\epsilon}_l}{2}\cdot \frac{1-{\sigma}_l}{2}\cdot {\epsilon}_1 \cdots {\epsilon}_{l-1}
\cdot {\sigma}_1\cdots {\sigma}_{l-1} \bigg)\, ;
$$
$$
A_k (\eta, \sigma)\, = \, \prod_{l=1}^k \bigg( \frac{1+{\eta}_l}{2}\, +\,
\frac{1-{\eta}_l}{2}\cdot \frac{1+{\sigma}_l}{2}\cdot {\eta}_1 \cdots {\eta}_{l-1}
\cdot {\sigma}_1\cdots {\sigma}_{l-1} \bigg)\, ;
$$
$$
N_k (\epsilon,\eta, \sigma)\, = \,A_k (\eta, \sigma)\,\cdot C_k (\epsilon, \eta \sigma)\,; 
$$
and where $\epsilon \sigma \, = \, {\epsilon}_1 \sigma_1, \, \dots, {\epsilon}_{k} \sigma_k$.

Proof. Immediately follows from the fact that

$$
{ {\x}}^{\frac{1-\epsilon}{2}}{\x}^{\frac{1-\sigma}{2}} \, = \,
C(\epsilon , \epsilon
\sigma)\cdot {\x}^{\frac{1-\epsilon \sigma}{2}}\, ;\;\qquad 
\bigg(\frac{\delta}{\delta \x}\bigg)^{\frac{1-\eta}{2}}{\x}^{\frac{1-\sigma}{2}} \, = \,
A(\eta, \eta \sigma)\cdot {\x}^{\frac{1-\eta \sigma}{2}}\,;
$$

which could be checked by induction on $k$.
%\bigskip

%$$
%\cases {(a^l \c) (\sigma) \, = \, \sigma_1 \cdots \sigma_{l-1}\,\cdot \frac {1+\sigma_l}{2} \cdot
%\c(\sigma_1,\dots ,\, - \sigma_l, \dots, \, \sigma_k)} \\
%{({a}^{*\,l} \c) (\sigma) \, = \, \sigma_1 \cdots \sigma_{l-1}\,\cdot \frac {1-\sigma_l}{2} \cdot
%\c(\sigma_1,\dots ,\, - \sigma_l, \dots, \, \sigma_k)} \endcases  \eqno{(4.6)}
%$$
%namely
%$$
%a^{\sharp \, l}\, = \, {(-1)}^{b_1^* b_1 + \dots + b_{l-1}^* b_{l-1}} b_l^{\sharp}\, , \eqno{(4.7)}
%$$
%where
%$$
%(b_l\c)(\sigma)\, = \,\frac {1+\sigma_l}{2} \cdot
%\c(\sigma_1,\dots ,\, - \sigma_l, \dots, \, \sigma_K)}\,,
%$$
%$$
%(b_l^*\c)(\sigma)\, = \,\frac {1-\sigma_l}{2} \cdot
%\c(\sigma_1,\dots ,\, - \sigma_l, \dots, \, \sigma_K)}\,,
%$$
%which is exactly the Jordan-Wigner transformation.

\bigskip
\noindent {\bf {Corollary}}. If $F(\x)\, = \, \sum_{\epsilon\, \in {\text{\bf{Z}}}_2^{\times k}} \,
f(\epsilon) \x^{\frac {1-\epsilon_l}{2}}\, \in \, {\text{\bf{G}}}(k)$, the image $F(a^*)$ 
of the linear operator $F({\x}) \, \in \, {\text{\bf {C}}}(2k)$ is
$$
(F(a^*) \c)(\sigma) \, = \, \sum_{\epsilon\, \in {\text{\bf{Z}}}_2^{\times k}} \, f(\epsilon)
C_k (\epsilon, \sigma) \c(\epsilon \sigma)\, , \eqno{(4.8)}
$$
and, in particular, when $F(\x) \, = \, \sum_{1\leq i < j \leq 2k} A_{ij} \x_i \x_j$
$$
\sum_{1\leq i < j \leq 2k} A_{ij}(a_i^* a_j^* \c)(\sigma)\, = 
$$
$$
= \sum_{(i,j)\in \Gamma(A)}  A_{ij} \frac {1-\sigma_i}{2} \frac {1-\sigma_j}{2}
\prod_{l=i+1}^{j-1} \sigma_l \c (\sigma_1,\dots , - \sigma_i, \dots , -\sigma_j, \dots ,
\sigma_{2k})\, ,\eqno{(4.9)}
$$
where $\Gamma(A) \, = \, \{(i,j), 1\leq i < j \leq 2k \; : A_{ij} \, \neq \, 0 \, \}$. 

\noindent Let $A$ be a linear operator acting on ${\Cal{H}}_k$ and $A(.,.)$ its matrix, given by
$$
(A\c)(\sigma) \, = \, \sum_{\sigma '\, \in {\text{\bf{Z}}}_2^{\times k}} A(\sigma,\sigma ')
\c (\sigma '). \eqno{(4.10)}
$$
For instance, from Proposition 2.1 we get that the matrix element of the operator 
$a^{*\frac{1-\epsilon}{2}} a^{\frac{1-\eta}{2}}$ is given by
$$
A(\sigma, \sigma ')\, = \, 
N_k(\epsilon, \eta, \sigma) \delta_{\epsilon \sigma , \eta \sigma '}  \,.
$$
Now from the formula (4.4) we obtain:
$$
\int^B \x^{\frac{1+\sigma}{2}} A \,\x^{\frac{1-\sigma '}{2}} {\D}_k \x \, = \, 
\triangle_k (\sigma) A(\sigma, \sigma ')\, , \qquad \forall A\, 
\in {\text{\bf {C}}}(2k)
\eqno{(E.1)}
$$
which relates the Berezin integral 
$\int^B \x^{\frac{1+\sigma}{2}} {\hat {A}} \,\x^{\frac{1-\sigma '}{2}} {\D}_k \x$
to the matrix $A(\sigma, \sigma ')$. This is our first basic formula.

We end the Section by some remarks about kernels of 
$A \, \in \, {\text{\bf {C}}}(2k) $. 
We recall that to each ${A}\,
\in \, {\text{\bf {C}}}(2k)$ corresponds a unique ${\text{Ker}}(A)(\x, \x ') \, \in \, {\text
{\bf{G}}}^2 (k)$,
the kernel of the linear operator ${A}$, such that 
$$
{{A}} F(\x) \, = \, \int^B {\text{Ker}}(A)(\x, \x ') F(\x ')\, {\D}_k \x '\; , 
\qquad \forall {F(\x)}\, \in {\text{\bf {G}}}(k).
\eqno{(4.12)}
$$
By exploiting Lemma 2.1, it is easy to see that for all positive 
integer $k$ the kernel ${\text{Ker}} ({\text{\bf {1}}}_{2k})(\x, \x ')$   
of the unit element ${{\text{\bf 1}}}_{2K}$ of ${\text{\bf {C}}}(2k)$ is
$$
{\text{Ker}} ({\text{\bf {1}}}_{2k})(\x, \x ')\, = \,\sum_{\sigma \, \in {\text{\bf{Z}}}_2^{\times k}} 
\triangle_k
(\sigma) \x^{\frac {1-\sigma}{2}} \x^{'\frac {1+\sigma}{2}}\, , \eqno{(4.13)}
$$
and more generally:
$$
{\text{Ker}} (A)(\x, \x ')\, = \, \sum_{\sigma, \sigma '\, \in {\text{\bf{Z}}}_2^{\times k}} 
\triangle_k
(\sigma ')A(\sigma, \sigma ') \x^{\frac {1-\sigma}{2}} \x^{'\frac {1+\sigma '}{2}}\, ,
 \eqno{(4.14)}
$$
which, together with 
$$
A(\sigma, \sigma ') \, = \, \triangle_k (\sigma) \iint^B
\x^{\frac {1+\sigma}{2}} {\text{Ker}} (A)(\x, \x ') \, \x^{'\frac {1-\sigma}{2}} d{\x '}_k
\cdots d{\x '}_1 d{\x}_k \cdots d{\x}_1\, , \eqno{(4.15)}
$$
relates the kernel ${\text{Ker}} (A)(\x, \x ')$ of ${{A}}$ and the matrix $A(\sigma, \sigma ')$.
\bigskip
{\bf{5. Semigroups Associated to Poisson Processes}}
\bigskip
We now turn to  semigroups of linear operators
acting on the hilbert space ${\Cal H}_{k}$.
We shall give a probabilistic representation of the semigroup ${\{ {\text {exp}}tL\}}_{t\geq 0}$ 
generated
by a non - trivial linear operator $L \, :\,{\Cal H}_{k}\, \to \, {\Cal H}_{k}$. It 
specializes the more general formulas introduced in [DJLS] to which 
we address the reader, nevertheless the Section will be self - 
contained.

We start with a special representation of  $L \, \in \,L({\Cal H}_{k},\, {\Cal H}_{k})$.
For all $\epsilon \, = \, (\epsilon_1 , \dots , \epsilon_k) \, \in \, {\text{\bf{Z}}}_2^{\times k},$ 
and different from the identity : $\epsilon \, \neq \, e_k \, = \, (1, \dots , 1)$, 
let $D^{\epsilon}$ be the self - adjoint 
difference operator:
$$
(D^{\epsilon} \c)(\sigma)\, = \, \c (\epsilon \sigma) \, - \, \c (\sigma)\, ; \eqno{(5.1)}
$$
which annihilates constants.

\noindent We observe that each $L \, \in \,L({\Cal H}_{k},\, {\Cal H}_{k})$ admits 
the representation
$$
L\, = \, \sum_{\epsilon \neq \e_k} {\lambda}_{\epsilon} (.) D^{\epsilon} \, - \, V(.) 
{\text{\bf {1}}}\, ,
$$

where the functions ${\lambda}_{\epsilon} (.), \, V(.) \, : {\text{\bf{Z}}}_2^{\times k}\, \to 
{\Bbb{C}}$
are related to the matrix $L(.,.)$ of the operator $L$ by formulas:

$$
{\text{i)}} \quad {\lambda}_{\epsilon} (\sigma)\, = \, L(\sigma,\epsilon \sigma); \qquad \qquad
{\text{ii)}}
\quad V(\sigma)\, = \, - \sum_{\sigma ' \in {\text{\bf{Z}}}_2^{\times k}} L(\sigma,\sigma ').
$$

Let $\Gamma(L)$ be the collection of $\epsilon \neq e_k$ such that $\lambda_{\epsilon} (.)$ is
not identically vanishing and $|\Gamma(L)|$ be its cardinality. If $\Gamma(L)\neq
\emptyset$ we call the operator L a difference operator. 

Now let ${\{ N_t^{\epsilon}\}}_{\epsilon \neq e_k}$ be a given collection of $(2^{k}-1)$
independent Poisson processes of unit parameter which we assume to be left - continuous.

%\noindent   i) $t \, \in \, [0, + \infty ) \, \to \,N_t^{\epsilon} $ is %left - continuous;

%\noindent  ii) $\,N_{t=0}^{\epsilon} \, = \, 0 \; a.s.$;

%\noindent iii) $\forall t, \, \triangle t>0, \, N_{t+\triangle %t}^{\epsilon} \, - \,
%N_t^{\epsilon}$ is independent of $N_s^{\epsilon} , \; 0\leq s \leq %t$, and
%$$
%{\text{Prob}}(N_{t+\triangle t}^{\epsilon} \, - \,N_t^{\epsilon} \, = \, %n) \, = \,
%\frac {e^{\triangle t} {(\triangle t)}^n}{n!}\, , \; n\, = \, 0, \, 1, \dots.
%$$

%\noindent  iv) $N_s^{\epsilon}$ and $N_{s'}^{\epsilon '}$ are %independent for $\epsilon \, 
%\neq \, \epsilon '$.

{\bf{Theorem 5.1}} (Probabilistic representation of semigroups)

\noindent Let  $L\, = \,  \sum_{\epsilon \in \Gamma(L)} {\lambda}_{\epsilon} (.) D^{\epsilon} \, - \,
 V(.) {\text{\bf {1}}}$ be a difference operator. Then
$$
(e^{tL}\c)(\sigma)\, = \,
$$
$$
= \, e^{t|\Gamma(L)|} {\text{\bf {E}}} \bigg( \c (\sigma{(-1)}^{N_t}) 
{\text{exp}} \{ \sum_{\epsilon \in \Gamma(L)}  \bigg( \int_{[0,t)} ln \, {\lambda}_{\epsilon}
(\sigma{(-1)}^{N_s}) d N_s^{\epsilon}\, - \,
$$
$$
-\, \int_0^t {\lambda}_{\epsilon} (\sigma{(-1)}^{N_s}) ds \bigg) \, - \, 
\int_0^t V(\sigma{(-1)}^{N_s}) ds\} \bigg)
$$
where $N_s \, = \, (N_s^1, \dots , N_s^k)$ with $N_s^l \, = \,\sum_{\epsilon \in \Gamma(L)}
\frac {1-{\epsilon}_l}{2} N_s^{\epsilon}$.

{\bf{Remark}} By convention ${\text{exp}}\int_{[0,t)} ln \, {\lambda}_{\epsilon}
(\sigma{(-1)}^{N_s}) d N_s^{\epsilon}$ vanishes if ${\lambda}_{\epsilon}
(\sigma{(-1)}^{N_s}) \, = \, 0$ for some $0<s<t$ such that $N_{s+}^{\epsilon} \, - 
N_s^{\epsilon}\, \neq \, 0$. We observe that it doesn't depend upon the choice of the branch 
of the logarithm.

Proof.  We follow the strategy explained in [DJLS]. 

\noindent i) The r.h.s. defines a semigroup 
${\{ P^t\}}_{t\geq0}$ of linear operators on ${\Cal H}_K$ by the Markov property of Poisson process. 
In order to complete the proof we must show that the infinitesimal generator of 
${\{ P^t\}}_{t\geq0}$ coincides with $L$. 

\noindent ii) Since each wave function $\c(.)$ is
a linear superposition of characters ${\chi}_n (\sigma)\, = \, {\sigma}^n \, = \, 
{\sigma}^{n_1}_{1}\dots {\sigma}^{n_k}_{k}, \; n \, = \, (n_1, \dots, n_k) \, \in \,
{({\text{\bf{Z}}}_2^*)}^{\times k} \, , \; {\text{\bf{Z}}}_2^* \, \equiv \, \{0,1\}$, we can only consider
the case of $\c(.) \, = \, {\chi}_n (.)$ for some $n \, \in {({\text{\bf{Z}}}_2^*)}^{\times k}$. 
Let ${\xi}_t^{\sigma, n}$ be the random variable 
$$
{\xi}_t^{\sigma, n} \, = \,\sum_{\epsilon \in \Gamma(L)} \int_{[0,t)} b_{\epsilon, n}
(\sigma{(-1)}^{N_s})  d N_s^{\epsilon} \, + \,\int_0^t  m (\sigma{(-1)}^{N_s}) ds\, ,
$$
where 
$$
b_{\epsilon, n} (\sigma)\, = \, i \pi \sum_{l=1}^k \frac {1-{\epsilon}_l}{2} n_l \, +
ln {\lambda}_{\epsilon}(\sigma)\, ;
$$
$$
m(\sigma) \, = \, - \sum_{\epsilon \in \Gamma(L)} {\lambda}_{\epsilon}(\sigma)\, - \, V(\sigma)\, .
$$
By definition, 
$$
(P^t {\chi}_n) (\sigma) \, = \, {\chi}_n (\sigma) e^{t|\Gamma (L)|} 
{\text{\bf E}}({\text{exp}}{\xi}_t^{\sigma, n})\, .
$$  

\noindent iii) The stochastic differential $d \, {\text{exp}}{\xi}_t^{\sigma, n}$ of the process 
$t \, \in \, [0, + \infty ) \, \to \, {\text{exp}}{\xi}_t^{\sigma, n}$ can be explicitly evaluated
as explained in [DJLS]. It turns out that 
$$ 
d \, {\text{exp}}{\xi}_t^{\sigma, n}\, = 
$$
$$
= \, ( {\text{exp}}{\xi}_t^{\sigma, n}) [ m (\sigma{(-1)}^{N_t}) dt \, +\, \sum_{\epsilon \in \Gamma(L)}
(e^{b_{\epsilon, n} (\sigma{(-1)}^{N_t})} -1)\, dN_t^{\epsilon}] \, =
$$
$$
= \, ( {\text{exp}}{\xi}_t^{\sigma, n}) [ -V(\sigma{(-1)}^{N_t}) dt \, +\, \sum_{\epsilon \in \Gamma(L)}
{\lambda}_{\epsilon}(\sigma{(-1)}^{N_t}) ({\epsilon}^n dN_t^{\epsilon} \, - \, dt) 
\, - \,\sum_{\epsilon \in \Gamma(L)} dN_t^{\epsilon}] \, .
$$
By taking the expectation of $d \, {\text{exp}}{\xi}_t^{\sigma, n}$ for $t=0$, since 
${\text{\bf E}}( dN_t^{\epsilon}) \, = \, dt$ and $(D^{\epsilon} {\chi}_n ) (\sigma)\, = \, 
({\epsilon}^n -1){\chi}_n ) (\sigma)$, it follows that $\frac {d}{dt} 
(P^t {\chi}_n)(\sigma) |_{t=0} \, = \, (L{\chi}_n ) (\sigma)$. Therefore the infinitesimal
generator of the semigroup ${\{ P^t\}}_{t\geq0}$ is exactly the difference operator $L$.
\bigskip
{\bf{Example}}  As an elementary illustration of Theorem 5.1, let $k=1$ and $(D \c)(\sigma) \, = \, 
\c (-\sigma) \, - \, \c(\sigma)$, then
$$
(e^{tD} \c )(\sigma) \, = \,{\text{\bf E}}(\c(\sigma{(-1)}^{N_t}) ) \, , \eqno{(5.2)}  
$$
where $N_t$ is the Poisson process with unit parameter. Indeed
$$
{\text{\bf E}}(\c(\sigma{(-1)}^{N_t}) )\, = \, \sum_{n=0}^{\infty} \frac {e^{-t} t^n}{n!} \c
(\sigma{(-1)}^n) \, = \, 
$$
$$
= \, e^{-t}\{ \c (\sigma){\text {cosh}} \,t \, + \,\c (-\sigma) {\text {sinh}} \, t \} \,
 = \, (e^{t D}\c)
(\sigma) \, .
$$

From Theorem 5.1 we obtain the matrix elements of the operator $e^{tL}$:
$$
e^{tL}(\sigma,\sigma ')\, = \,
$$
$$
= \, e^{t|\Gamma(L)|} {\text{\bf {E}}} \bigg( \prod_{l=1}^k \frac {1+ {\sigma}_l {\sigma '}_l{(-1)}^{N_t^l}}
{2} 
{\text {exp}} \{ \sum_{\epsilon \in \Gamma(L)}  \bigg( \int_{[0,t)} ln \, {\lambda}_{\epsilon}
(\sigma{(-1)}^{N_s}) d N_s^{\epsilon}\, - \,
$$
$$
-\, \int_0^t {\lambda}_{\epsilon} (\sigma{(-1)}^{N_s}) ds \bigg) \, - \, 
\int_0^t V(\sigma{(-1)}^{N_s}) ds\} \bigg)\, , \eqno{(E.2)}
$$ 
which is the second important equality and will be used in the next Section evaluating Berezin integrals.

\bigskip
{\bf{6. Berezin Integrals and Poisson Processes in the $\s$ - Representation}}
\bigskip

We shall consider Berezin integrals of the form:  
$$
\int^B {\x}^n {\text {exp}} (-S(\x)) {\D}_k \x \, = \, \int^B {\x}_1^{n_1}{\x}_2^{n_2} \cdots
{\x}_k^{n_k} {\text {exp}}  (-S(\x)) {\D}_k \x\, , \eqno{(6.1)}
$$
where  $n \, = \, (n_1, \dots, n_k) \, \in \,{({\text{\bf{Z}}}_2^*)}^{\times k}$ and $S(\x) \neq 
c\cdot {\text {\bf 1}}_k$ is a non - trivial element of {\G}$(k)$. It could be interpreted as 
the ``action''  for the (Euclidean) ``Fermi field'' $\{ {\x}_1, \dots ,{\x}_k \}$ in which case 
(6.1) would provide all (unnormalized) ``correlation functions'' or (Euclidean) ``Green
functions'' of the field. In particular, for $n\, = \, (0, \, 0, \dots , 0)$, (6.1) gives the
``partition function'':
$$
Z[S] \, = \, \int^B {\text {exp}} (-S(\x)){\D}_k \x\, . \eqno{(6.2)}
$$
{\bf{Remark}}. We could easily consider more general integrals by taking
for $S$ an element of {\bf{C}}(2k). Integrals of the form (6.1) cover
a large class of physical applications. 

Let $s(.) \, : {\text{\bf{Z}}}_2^{\times k} \, \to \,{{\Bbb{C}}} $ be defined as
$$
s(\e)\, = \, {\triangle}_k (\e) \int^B {\x}^{\frac {1+\e}{2}}S(\x) {\D}_k \x\, . \eqno{(6.3)}
$$
with $\triangle_k (\e)$ as in (4.5).

\noindent {\bf {Theorem 6.1}} (Berezin integrals as Poisson averages)
For each $k=1, \, 2, \, \dots,$ each non - trivial $S(\x)\in {\text{\G}}_k$ and for all 
$n=(n_1, \dots,
n_k) \in  (\text{\bf{Z}}_2^*)^{\times k}$
$$
\int^B {\x}_1^{n_1}\cdots{\x}_k^{n_k}{\text {exp}} (-S(\x)){\D}_k \x\, =
$$
$$
= \,{\triangle}_k (-{(-1)}^n)e^{(|\Ga(S)|-s(e_k))} {\text {\bf{E}}}
\bigg( \prod_{l=1}^k \frac {1- {(-1)}^{N_1^l+n_l}}{2}\, \times \eqno{(E.3)}
$$
$$
\times \, {\text {exp}}\{\sum_{\e \in \Ga(S)}
\int_{[0,1)} ln(-s(\e)C_k(\e, -{(-1)}^{N_s+n}))dN_s^{\epsilon}\}\bigg) \,,
$$
where $s(\e)$  is given by (6.3).

In particular
$$
Z[S]\, = \, e^{(|\Ga(S)|-s(e_k))} {\text {\bf{E}}}
\bigg( \prod_{l=1}^k \frac {1- {(-1)}^{N_1^l}}{2}\, \times
$$
$$
\times \,{\text {exp}}\{\sum_{\e \in \Ga(S)}
\int_{[0,1)} ln(-s(\e)C_k(\e, -{(-1)}^{N_s}))dN_s^{\e}\} \bigg)\,. \eqno{(6.5)}
$$

Proof.  
Using formula (4.4) we get:
$$
S(\x) \, = \, \sum_{\e \in {\text{\bf{Z}}}_2^{\times k}} s(\e){\x}^{\frac {1-\e}{2}}\, ,  \eqno{(6.6)}
$$
and, by hypothesis, the subset $\Ga(S) \, = \, \{\e \in {\text{\bf{Z}}}_2^{\times k} , \e \neq e_k\,:
s(\e)\neq0\} $ is non empty. To the operator $-S({\hat {\x}}) \in {\text{\bf {C}}}(2k)$ corresponds the
image $L\, = \, -S(a^*) \in L(\H_k, \H_k)$ which is the difference operator:
$$
L \, = \, \sum_{\e \in \Ga(S)} {\l}_{\e}(\cdot) D^{\e} \, - \, V(.) {\text{\bf {1}}}\, ,  \eqno{(6.7)}
$$
with 
$$
{\l}_{\e}(\s)\, = \, - s(\e)C_k (\e,\s)\, , \eqno{(6.8)}
$$
$$
V(\s)\, = \, \sum_{\e \in {\text{\bf{Z}}}_2^{\times k}} s(\e)C_k (\e,\s)\,=\, s(e_k)\, - \,
\sum_{\e \in \Ga(S)} {\l}_{\e}(\s)\, .\eqno{(6.9)}
$$

From equations (E.1), (E.2)  
and equality  
$$
-V(\s) \, - \, \sum_{\e \in \Ga(S)} {\l}_{\e}(\s)\, = \, -s(e_k)\, ,
$$
follows that
$$
-\int_0^1 V(-{(-1)}^{N_s +n})ds \, - \, \sum_{\e \in \Ga(S)} \int_0^1  {\l}_{\e}
(-{(-1)}^{N_s +n})ds \, = \, -s(e_k)\, .
$$
\bigskip

$C_k (\e, \s)$ is defined in Proposition $4.1$ and $N_s, N_s^l$
in Theorem $5.1$.

We now make an important remark.

{\bf{Remark }}. In equation (E.3) only
the trajectories of the Poisson process with zero or one jump
contribute to the expectation. In fact as soon as one of the factors
$(1+{(-1)}^{N_t^r + n_r})$ in $C_k$ vanishes the stochastic integral equals
$-\infty$. On the other hand $N_t^r$ is a sum of independent processes and the event of two processes jumping at the same instant has zero probability.

We now take advantage of the fact that the calculation of a Berezin 
integral has been reduced to an ordinary  integral and we derive
simple estimates. From Theorem (6.1) we get
$$
\bigg|\int^B {\x}_1^{n_1}\cdots{\x}_k^{n_k}{\text {exp}} (-S(\x)){\D}_k \x\bigg| \, \leq
$$
$$
\leq \, e^{(|\Ga(S)|- Re s(e_k))} {\text {\bf{E}}}
\bigg( \prod_{l=1}^k \frac {1- {(-1)}^{N_1^l+n_l}}{2}\, \times
$$
$$
\times \,{\text {exp}}\{\sum_{\e \in \Ga(S)}
\int_{[0,1)} ln(|s(\e)| \bigg|C_k(\e, -{(-1)}^{N_s + n})\bigg|) 
dN_s^{\epsilon}\} \, \leq  \eqno{(6.10)}
$$
$$
\leq \, e^{(|\Ga(S)|- Re s(e_k))} {\text {\bf{E}}}
\bigg(\chi_{\{N_1^{\epsilon}=0,1\}} \prod_{l=1}^k \frac {1- {(-1)}^{N_1^l+n_l}}{2} \prod_{\e \in \Ga(S)} {|s(\e)|}^{N_1^{\e}}
\bigg)\, = \,
$$
$$
e^{(- Re s(e_k))}\frac{1}{2^k} \sum_{\rho_1,....\rho_k=0,1}(-1)^{\Sigma_{i=1}^k\rho_i(1+n_i)}
\times \, \prod_{\epsilon\in\Gamma(S)}\{1 + (-1)^{\Sigma_{i=1}^k\rho_i
\frac{(1-\epsilon_i)}
{2}} |s(\epsilon)| \} \, 
$$
since $|C_k (\e, \s)| \, \leq \, 1$ and $\int_{[0,1)} d N_s^{\e} \, = \, 
N_1^{\e}$; $\chi$ is the characteristic function of the event indicated.

\bigskip
{\bf{7. Gaussian Berezin Integrals}}
\bigskip

Gaussian Berezin integrals are a particular case when the ``action'' $S$ is  bilinear 
in the ``Fermi field''. Let {\G}$^2 (k)$ be the Grassmann algebra generated by
$\{{\bar {\eta}}_1,\dots,{\bar {\eta}}_k; \eta_1,\dots, \eta_k \} $ and $(B)_{ij}$ any
$k \times k$ matrix. Let us consider the integral
$$
\int^B {\bar {\eta}}_1^{{\bar {\nu}}_1}\cdots{\bar {\eta}}_k^{{\bar {\nu}}_k}
{\eta}_1^{{\nu}_1} \cdots{\eta}_k^{{\nu}_k}{\text {exp}} \{ - \frac {1}{2} 
\sum_{i,j=1}^k
B_{ij}{\bar {\eta}}_i \eta_j\} \D_k \eta \D_k {\bar {\eta}} \, , \eqno{(7.1)}
$$
where ${{\bar {\nu}}_l}, {{\nu}_l} \, \in {\text{\bf{Z}}}_2^* \, = \, 
\{ 0,1\}.$  Making the substitution  $\x_1 = {\bar {\eta}}_1, \dots , 
\x_k = {\bar {\eta}}_k; \;
\x_{k+1}= \eta_1, \dots, \x_{2k}=\eta_k, $ we get 
$$
\sum_{i,j=1}^k B_{ij}{\bar {\eta}}_i \eta_j\, = \, 
\sum_{r,s=1}^{2k} A_{rs} {\x}_r \x_s \, ,
$$
where $(A_{rs})$ is the $2k \times 2k$ antisymmetric matrix defined by
$$
A_{rs} \, = \, \cases {\; \; 2^{-1}B_{r(s-k)} \; {\text{when}}\; r=1, \dots,k \; {\text{and}}\;
s= k+1, \dots, 2k };\\
{-2^{-1}B_{s(r-k)} \; {\text{when}}\; r=k+1, \dots,2k \; {\text{and}}\;
s= 1, \dots, k };\\
{\; \; 0 \qquad {\text{otherwise}}}. \endcases  \eqno{(7.2)} 
$$
Therefore the Gaussian integral (7.1) can be written in the following form:
$$
\int^B {\bar {\eta}}_1^{{\bar {\nu}}_1}\cdots{\bar {\eta}}_k^{{\bar {\nu}}_k}
{\eta}_1^{{\nu}_1} \cdots{\eta}_k^{{\nu}_k}{\text {exp}} \{ - \frac {1}{2} \sum_{i,j=1}^k
B_{ij}{\bar {\eta}}_i \eta_j\} \D_k \eta \D_k {\bar {\eta}} \,=
$$
$$
=\; \int^B {\x}_1^{n_1}\cdots{\x}_{2k}^{n_k}{\text {exp}} (-\frac {1}{2}
\sum_{r,s=1}^{2k} A_{rs} {\x}_r \x_s ){\D}_{2k} \x \, , \eqno{(7.3)}
$$
and from now on we use the last form. Let $\Ga(A)$ be the set:
$$\Ga(A) \, = \, \{(r,s), \; 1\leq r < s \leq 2k \; :  A_{rs} \neq 0\}\, , \eqno{(7.4)}
$$
which we suppose not to be empty.  From (4.9)  we get that the generator $L$  of the corresponding
semigroup is given by
$$
(L\psi)(\s) \, = \, 
$$
$$
= \, \sum_{(r,s) \in \Ga(A)} A_{rs} \frac{1-{\s}_r}{2}\frac{1-{\s}_s}{2} 
\prod_{l=r}^{s-1} {\s}_l \psi (\s_1, \dots, -\s_r, \dots, -\s_s, \dots, \s_{2k})\, , \eqno{(7.5)}
$$
and therefore

$$\int^B {\x}_1^{n_1}\cdots{\x}_{2k}^{n_{2k}}{\text {exp}} (-\frac {1}{2}
\sum_{r,s=1}^{2k} A_{rs} {\x}_r \x_s ){\D}_{2k} \x \, =
$$
$$
= \,{\triangle}_k (-{(-1)}^n)e^{|\Ga(A)|} {\text {\bf{E}}}
\bigg( \prod_{l=1}^{2k} \frac {1- {(-1)}^{N_1^l+n_l}}{2}\, \prod_{(r,s) \in \Ga(A)} 
{(A_{rs})}^{N_1^{(r,s)}} \times \eqno{(E.4)}
$$
$$
\times \, {\text {exp}}\{\sum_{(r,s) \in \Ga(A)}
\int_{[0,1)} ln \bigg[ \frac {(1+{(-1)}^{N_t^r + n_r})(1+{(-1)}^{N_t^s + n_s})}{4}
\prod_{l=r}^{s-1} {(-1)}^{s-r} {(-1)}^{N_t^l + n_l}\bigg] dN_t^{(r,s)} \} \bigg)\, ,  
$$
where ${\{ N_t^{(r,s)} \}}_{(r,s) \in \Ga(A)}$ is a family of independent Poisson
processes with unit parameter and
$$
N_t^l \, = \, \sum_{(l,m) \in \Ga(A)} N_t^{(l,m)}\, + \, \sum_{(m,l) \in \Ga(A)} N_t^{(m,l)}
\quad {\text {for all}} \; 1\leq l \leq 2k.
$$

\noindent {\bf{Example.}} Let $k=1, \, n = (0,0)$ and $(A_{rs}) = i\s_2 $ (Pauli matrix), then
$$
-1 \, = \, \int^B e^{-{\bar {\eta}} \eta} d\eta d{\bar {\eta}} \, = \,
\int^B e^{- \frac {1}{2} {\Sigma_{r,s=1}^2} A_{rs}\x_r \x_s }\D_2 \x \, =
$$
$$
= \, e{\text {\bf {E}}}\bigg( \frac {1- {(-1)}^{N_1}}{2} {\text { {exp}}}
\int_{[0, 1 )} ln \bigg[ - {(-1)}^{N_t} \frac {1+ {(-1)}^{N_t}}{2} 
\bigg]dN_t \bigg) \, \eqno{(7.6)}   
$$
This equality can be checked directly:
indeed the random variable under expectation vanishes unless the 
Poisson process has exactly one jump for some $0<s<1$, in which case $N_1=1$, and which happens
with probability $e^{-1}$.

{\bf{Remark }}. Suppose we want to use our representation
of Gaussian Berezin integrals in a numerical simulation.
Let $2k$ be the total number of $\eta$ and $\bar{\eta}$. The
generic case will require the generation of $k(2k-1)$ independent
Poisson processes. For a Dirac
field on a finite d-dimensional lattice $\Lambda$ the maximum number
of Poisson processes involved
is $L|\Lambda|(2L|\Lambda|-1))$ where $L$ is the number
of components of the field. This is a large number.
However if we want to calculate for example the propagator of the free field
we need much less. In fact                                         
the matrix $B_{ij}$ which is the lattice version
of the differential operator ${\partial}\!\!\!/+{A}\!\!\!/-M$
couples only nearest neighbours. Therefore the required number of independent Poisson processes is reduced to roughly
$dL|\Lambda|$ since each site has $2d$ nearest neighbours.

\bigskip
\bigskip
\noindent{\bf{Acknowledgments.}}  We would like to thank M. Cassandro 
and M.E. Vares for useful discussions. We thank also an anonymous referee
for pointing out to us some relevant references. 
One of the authors (V.S.) 
would like to thank I.N.F.N. and the Universit\`a  La Sapienza, for support and hospitality.
\bigskip
\bigskip
\noindent{\bf{Appendix 1.}}
\bigskip

{\bf {Proposition A.1.}}
The kernel of the operator $\g^{(\nu,x,\mu)}$ given by (3.8) and acting on {\bf{G}}$(k)$ is given by
the following formulas:
$$
{\text {Ker}}\g^{(\nu,x,\mu)}(\xi,\x') = \int^B \prod_{j=1}^n(\xi_{\nu_j} +
i\rho_{\nu_j})\cdot
\prod_{j=1}^{\ell}(2\xi_{x_j}\,\rho_{x_j}+i{1})\cdot
$$
$$
\cdot \prod_{j=1}^m \frac{1}{i} (\xi_{\mu_j} - i\rho_{\mu_j})\cdot 
\exp\big\{-i\sum_{r=1}^k \rho_r(\xi_r-\x'_r)\big\}d\rho;   \eqno{(A.1)}
$$
if $k$ is even, and
$$
{\text {Ker}}\g^{(\nu,x,\mu)}(\xi,\x') = \int^B \prod_{j=1}^n(\xi_{\nu_j} +
\rho_{\nu_j})\cdot \prod_{j=1}^{\ell}({1}-
2\xi_{x_j}\,\rho_{x_j})\cdot
$$
$$
\cdot \prod_{j=1}^m \frac{1}{i} (\xi_{\mu_j} - \rho_{\mu_j})\cdot \exp\big\{-
\sum_{r=1}^n \rho_r(\xi_r+\x'_r)\big\}d\rho;   \eqno{(A.2)}
$$
if $k$ is odd.

\noindent  The proof is straightforward, and we omit it.

{\bf{Remark.}} Here we expressed elements of the algebra {\bf{G}}$_0 (k)$ in so 
called ``Fourier-transform'' form. In fact we
just multiply initial expression by $\rho$-monomials and then integrate over $\rho$.

Let us set:
$$
L = \sum_{(\nu,x,\mu)} h_{(\nu,x,\mu)}\,\g^{(\nu,x,\mu)}\,.
$$
where $\g^{(\nu,x,\mu)} \, \in \text{\bf{C}}(2k)$, and are of the form (3.8), 
and $h_{(\nu,x,\mu)}\, \in {\Bbb{C}}$.

After computations using (A.1) and (A.2) we get:
$$
\text{{Ker}}(e^{\frac 1n\,L})(\xi,\x') = \text{{Ker}}({1} - \frac 1n L)(\xi,\x') + 
o(\frac 1n)\, =
$$
$$
= \int^B \exp\big\{-\frac 1n \sum_{(\nu,x,\mu)} h_{(\nu,x,\mu)} \prod_{\nu}
(\xi_\nu + i\rho_\nu) \prod_x (2\xi_x\rho_x + i{1}) \cdot
$$
$$
\cdot \prod_{\mu} \frac 1i (\xi_\mu-i\rho_\mu) - \sum_{r=1}^n
\rho_r(\xi_r+\x'_r)\big\}d\rho_n \cdot\ldots\cdot d\rho_1 + o(\frac 1n),
$$
for $k$ even and analogous formula holds for $k$ being odd.

{\bf{Remark.}} Since n is fixed we always understand convergence as a pointwise
convergence in a finite dimensional linear space.

Finally applying Trotter's formula we have:
$$
e^{tL}\,f(\xi) =
$$
$$
\lim_{m\to\infty} \int^B\ldots\int^B \text{{Ker}}(e^{\frac
tm\,L})\left[\xi,\x\left(\frac{(m-1)t}{m}\right)\right]\cdot
$$
$$
\cdot \text{{Ker}} (e^{\frac tm\,L})\left[\x\left(\frac{(m-1)t}{m}\right), \x\left(\frac{(m-
2)t}{m}\right)\right]\cdot \ldots
$$
$$
\cdot \ldots \text{{Ker}} (e^{-\frac tm\,L})\left[\x\left(\frac tm \right),\x'\right]
f(\x)d\x'\,d\x\left(\frac tm \right) \cdot\ldots\cdot d\x\left(\frac{(m-1)t}{m}\right) =
$$
$$
=\lim_{m\to\infty} \int^B\ldots\int^B \exp\bigg\{\sum_{j=1}^m \bigg(-
\frac{j\cdot t}{m}\cdot\sum_{(\nu,x,\mu)} h_{(\nu,x,\mu)} \prod_{\nu}
(\xi_\nu+i\rho_\nu)\cdot 
$$
$$
\cdot \prod_x (2\xi_x\rho_x + i{1})\cdot \prod_{\mu} \frac 1i (\xi_\mu -
 i\rho_\mu)\bigg) -
$$
$$
-\sum_{j=1}^m \sum_{r=1}^n \rho_r
\left(\frac{j\cdot t}{m}\right)\left(\x_r\left(\frac{j\cdot t}{m}\right) +
\x_r\left(\frac{(j-1)t}{m}\right)\right)\bigg\}f(\x')\cdot
$$
$$
\cdot d\rho\left(\frac tm\right) \cdot\ldots\cdot
d\rho(t)d\x'd\x\left(\frac tm\right) \cdot\ldots\cdot d\x(t) =
$$
$$
= \lim_{m\to\infty} \int^B Q_t^m(\xi,\x')f(\x')d\x'; \eqno{(A.3)}
$$
and from which we get Theorem 3.1. 
\bigskip
\noindent{\bf{References.}}
\bigskip
\noindent [B1]   Berezin, F. {\sl The Method of  Second Quantisation}, Academic
Press Inc., 1966.

\noindent [B2]  Berezin, F. {\sl The Planar Ising Model}, Russian Math. Surveys
{\bf 24}, n. 3, (1969).

\noindent [B3] Berezin, F. {\sl On the Number of Planar Non-self-intersecting
Contours on a Planar Lattice}, Math. USSR Sbornik {\bf 88}, 269, (1972).

\noindent [BM] Berezin, F., Marinov M. S. {\sl Particle Spin Dynamics},
Ann. Phys. {\bf 104}, 336, (1977).

\noindent [BR]  Bratteli, O., Robinson, D. W. {\sl Operator algebras and 
quantum statistical mechanics II},
Springer Verlag,  1981.

\noindent [BZ] Bismut, J.-M., Zhang, W., {\sl An Extension of a Theorem
by Cheeger and M\"uller}, Asterisque 205, 1992.

\noindent [DJLS] De Angelis, F., Jona-Lasinio, G., Sirugue, M. 
{\sl Probabilistic
Solution of Pauli Type Equations}  J.Phys.A: Math.Gen. {\bf 16}, 2433, (1983).

\noindent [D]   De Witt, B. {\sl Supermanifolds}, Cambridge Univ. Press,  1984.

\noindent [ER] Erd\"os, L, {\sl Gaussian Decay of the Magnetic Eigenfunctions},
Geom. Func. Anal. {\bf 6}, 231, (1996). 

\noindent [F]  Fyodorov, Y. V., {\sl Basic
Features of Efetov's Supersymmetry Approach} in {\sl Mesoscopic Quantum
Physics}, Les Houches 1994, Elsevier 1995.

\noindent [FS] Faddeev, L. D., Slavnov, A. A., {\sl Gauge Fields},
Benjamin/Cummings, 1980.

\noindent [MA] Marinov, M. S.; {\sl Path Integrals in Quantum Theory},
Phys. Reports {\bf 60}, 1, (1980).

\noindent [M]   Malyshev, V., Minlos, R.; {\sl Gibbs Random Fields}, 
Kluwer, 1991.

\noindent [MIS] Malyshev, V., Ignatiuk, I., Sidoravicius, V.; {\sl The 
Convergence
of the Method of the Second Quantization II}, SIAM Prob. Theory and Appl.  
(1993).

\noindent [RZ] Regge, T., Zecchina, R.; {\sl The Ising Model on Group Lattices
of Genus $>$ 1}, J. Math. Phys. {\bf 37}, 2796, (1996).

\noindent [R]   Rogers, A.; {\sl Fermionic Path Integration and 
Grassmann Brownian Motion}, Commun. Math. Phys. {\bf 113}, 353, (1986).

\noindent [S]   Seiler, E., {\sl  Gauge Theories as a Problem of Constructive
Field Theory and Statistical Mechanics}, Lecture Notes in Phys. 159,
Springer Verlag, 1982.

\noindent [VWZ] Verbaarschot, J. J. M., Weidenmuller, H. A., 
Zirnbauer, M.R. 
{\sl Grassmann Integration
in Quantum Physics: the Case of Compound - Nucleus Scattering},
 Physics Reports {\bf 129}, No 6, 367, (1985). 

\noindent [W] Wigner, E. P., {\sl Group Theory}, Academic Press, 1959.

\noindent [Z] Zirnbauer, M. R., {\sl Riemaniann
Symmetric Superspaces and their Origin in Random Matrix Theory}, J. Math. Phys.
{\bf 37}, 4986 (1996).

%\printindex

\end{document}